\documentclass[aps,preprint,amsmath,amssymb,superscriptaddress,nofootinbib]{revtex4}
\usepackage{multirow}
\usepackage{graphicx}
\usepackage{mathtools}

\begin{document}

\title{Model-independent study on the anomalous $\tau \bar{\tau}\gamma$ couplings at the ILC}

\author{M. K\"{o}ksal}
\email[]{mkoksal@cumhuriyet.edu.tr} \affiliation{Department of
Optical Engineering, Cumhuriyet University, 58140, Sivas, Turkey}

\begin{abstract}

The potential of the process $\gamma \gamma \rightarrow \tau \bar{\tau}\gamma$  is examined in a model-independent way using the effective Lagrangian approach for the ILC which is designed with standard configurations of 0.25 TeV/2000 fb$^{-1}$, 0.35 TeV/200 fb$^{-1}$ and 0.5 TeV/4000 fb$^{-1}$. The limits obtained for $\tilde{a}_{\tau}$ and $\tilde{{d}_{\tau}}$ parameters defining the anomalous $\tau \bar{\tau}\gamma$ couplings at $95\%$ confidence level and systematic uncertainties of $\delta_{sys}=0,5,10 \%$ are compared with the experimental results. The best limits obtained without systematic error on the anomalous couplings are $-0.00082<\tilde{a}_{\tau}<0.00050$ and $|\tilde{{d}_{\tau}}|<3.5969 \times 10^{-18}$ $e\,cm$, respectively.
Thus, our results show that $\gamma \gamma$ collisions at the ILC lead to a remarkable improvement in the existing experimental limits on the anomalous magnetic and electric dipole moments of the tau lepton.
\end{abstract}

\maketitle

\section{Introduction}

New physics beyond the Standard Model (SM) presents the theoretical developments needed to enlighten the lacks of the SM, such as the strong CP problem, neutrino oscillations, matter$-$antimatter asymmetry in the universe. One of the ways to research new physics beyond the SM is the effective Lagrangian method.
The effective Lagrangian method is based upon the assumption that at higher energy regions beyond the SM, there is a more fundamental physics that reduces to the SM at lower energy regions. In this method, one adds higher dimensional effective operators suppressed by an energy cut-off ($\Lambda$) with the SM fields and obtain the interactions after symmetry breaking. Therefore, the new physics contributions on $\tau\bar{\tau}\gamma$ interactions through the effective Lagrangian method can be examined.

In this work, we study the effects of the anomalous $\tau\bar{\tau}\gamma$ couplings defined with the effective Lagrangian method in the model-independent approach between the tau lepton and the photon for the process $\gamma \gamma \rightarrow \tau \bar{\tau}\gamma$ at the International Linear Collider (ILC).
The most general anomalous vertex function determining $\tau\bar{\tau}\gamma$ interaction between two on-shell the tau lepton and a photon is given by \cite{d25,d26}

\begin{eqnarray}
\Gamma^{\nu}=F_{1}(q^{2})\gamma^{\nu}+\frac{i}{2 m_{\tau}}F_{2}(q^{2}) \sigma^{\nu\mu}q_{\mu}+\frac{1}{2 m_{\tau}}F_{3}(q^{2}) \sigma^{\nu\mu}\gamma^{5}q_{\mu}.
\end{eqnarray}
Here, $\sigma^{\nu\mu}=\frac{i}{2}(\gamma^{\nu}\gamma^{\mu}-\gamma^{\mu}\gamma^{\nu})$, $q$ represents the momentum transfer to the photon and $m_{\tau}=1.777$ GeV shows the tau lepton's mass. $F_{1}(q^{2})$ and $F_{2}(q^{2})$ are the Dirac and Pauli form factors, $F_{3}(q^{2})$ is the electric dipole form factor. The last term $\sigma^{\nu\mu}\gamma^{5}$ breaks the CP symmetry, so the coefficient $F_{3}(q^{2})$ determines the strength of a possible CP violation process, which might originate from new physics. $F_{1}(q^{2}),F_{2}(q^{2})$ and $F_{3}(q^{2})$ form factors in limit $q^{2} \rightarrow 0$ are equal to the formulas below

\begin{eqnarray}
F_{1}(0)=1,\: F_{2}(0)=a_{\tau},\: F_{3}(0)=\frac{2m_{\tau}d_{\tau}}{e},
\end{eqnarray}
where ${a}_\tau$ is the magnetic dipole moment of the tau lepton and ${d}_\tau$ is the electric dipole moment of the tau lepton.

In a lot of works examining the anomalous magnetic and electric dipole moments of the tau lepton, the tau leptons or the photon in $\tau\bar{\tau}\gamma$ couplings are off-shell. In this case, since the tau lepton is off-shell, the couplings analyzed in those works are not the anomalous $a_{\tau}$ and $d_{\tau}$. Hence, we will name the anomalous magnetic and electric dipole moments of the tau lepton examined as $\tilde{a}_{\tau}$ and $\tilde{d}_{\tau}$ instead of $a_{\tau}$ and $d_{\tau}$. Thus, the possible deviation from the SM predictions of $\tau\bar{\tau}\gamma$ couplings could be examined in the effective Lagrangian method. In this method, the anomalous $\tau\bar{\tau}\gamma$ couplings are parameterized using high dimensional effective operators. In this work, we consider the dimension-six operators that contribute to the magnetic and electric dipole moments of the tau lepton. These operators are presented as follows \cite{al12}

\begin{eqnarray}
Q_{LW}^{33}=(\bar{\ell_{\tau}}\sigma^{\mu\nu}\tau_{R})\sigma^{I}\varphi W_{\mu\nu}^{I},
\end{eqnarray}

\begin{eqnarray}
Q_{LB}^{33}=(\bar{\ell_{\tau}}\sigma^{\mu\nu}\tau_{R})\varphi B_{\mu\nu}.
\end{eqnarray}
Here, $\varphi$ and $\ell_{\tau}$ represent the Higgs and the left-handed $SU(2)$ doublets, $\sigma^{I}$ show the Pauli
matrices and $W_{\mu\nu}^{I}$ and $B_{\mu\nu} $ are the gauge field strength tensors. Thus, the effective Lagrangian can be written as follows

\begin{eqnarray}
L_{eff}=\frac{1}{\Lambda^{2}} [C_{LW}^{33} Q_{LW}^{33}+C_{LB}^{33} Q_{LB}^{33}+h.c.].
\end{eqnarray}

\noindent After the electroweak symmetry breaking, contributions to the magnetic and electric dipole moments of the tau lepton are given by

\begin{eqnarray}
\kappa=\frac{2 m_{\tau}}{e} \frac{\sqrt{2}\upsilon}{\Lambda^{2}} Re[\cos\theta _{W} C_{LB}^{33}- \sin\theta _{W} C_{LW}^{33}],
\end{eqnarray}

\begin{eqnarray}
\tilde{\kappa}=-\frac{\sqrt{2}\upsilon}{\Lambda^{2}} Im[\cos\theta _{W} C_{LB}^{33}- \sin\theta _{W} C_{LW}^{33}].
\end{eqnarray}
Here, $\upsilon$ represents the vacuum expectation value and $\sin\theta _{W}$ shows the weak mixing angle.
The relations between $\kappa$ and $\tilde{\kappa}$ parameters with $\tilde{a}_\tau$ and $\tilde{d}_\tau$ are defined by

\begin{eqnarray}
\kappa=\tilde{a}_\tau, \;\;\;\; \tilde{\kappa}=\frac{2m_\tau}{e}\tilde{d}_\tau.
\end{eqnarray}

The electron and muon anomalous magnetic moments can be studied with high sensitivity via spin precession experiments. On the other hand, since the tau lepton has a much shorter lifetime than other leptons it is extremely difficult to measure the magnetic moment of the tau lepton by using spin precession experiments. Instead of spin precession experiments, the magnetic moment measurement of the tau lepton is carried out at the collider experiments. Besides, new physics beyond the SM is anticipated to modify the SM prediction of the anomalous magnetic moment of a lepton $\ell$ of $m_{\ell}$ mass by a contribution of order $\sim m_{\ell}^{2}/\Lambda^{2}$. Thus, given the large factor $(m_{\tau}/m_{\mu})^{2}\cong 283$, the anomalous magnetic moment of the tau lepton is much more sensitive than the one of the muon to the electroweak and new physics effects which give contribution $\sim m_{\ell}^{2}$, making its measurement an excellent opportunity to unveil or constrain new physics effects.

Experimental limits at $95\%$ confidence level on the magnetic moment of the tau lepton were derived the processes $e^{-}e^{+}\rightarrow \tau^{-} \tau^{+} \gamma$ and $e^{-}e^{+}\rightarrow e^{-}\gamma^{*} \gamma ^{*}e^{+} \rightarrow e^{-} \tau^{-} \tau^{+} e^{+}$ by L3, OPAL, DELPHI and BELLE Collaborations by the Large Electron Positron Collider (LEP)\cite{l3,op,de}

\begin{eqnarray}
\text{L3}:-0.052<\tilde{a}_{\tau}<0.058,
\end{eqnarray}
\begin{eqnarray}
\text{OPAL}: -0.068<\tilde{a}_{\tau}<0.065,
\end{eqnarray}
\begin{eqnarray}
\text{DELPHI}:-0.052<\tilde{a}_{\tau}<0.013.
\end{eqnarray}

In the interaction between the tau lepton and photon, another contribution is the effect that violates the CP that generates the electric dipole moment. The SM is not sufficient to understand the source of the CP violation. In the SM, there is no CP violating interactions in leptonic interactions but it is probable that multi-loop contributions from the quark sector implicitly cause CP violation that is too small to detect  \cite{4,5}. The electric dipole moment of the tau lepton occurs only at three-loop in the SM and is thus extremely suppressed. If there considers a coupling of leptons in new physics beyond the SM, electric dipole moment may induce the detectable size of CP-violation \cite{11,12,13,bok,yam,yam1,yamm,yamm1}.

The experimental results on the $\tilde{d}_{\tau}$ coupling at $95\%$ confidence level at the LEP are \cite{l3,op,de,d1}

\begin{eqnarray}
\text{L3}:|\tilde{d}_{\tau}|<3.1 \times 10^{-16}\, e\,cm,
\end{eqnarray}
\begin{eqnarray}
\text{OPAL}:|\tilde{d}_{\tau}|<3.7 \times 10^{-16}\, e\,cm,
\end{eqnarray}
\begin{eqnarray}
\text{DELPHI}:|\tilde{d}_{\tau}|<3.7 \times 10^{-16}\, e\,cm,
\end{eqnarray}
\begin{eqnarray}
\text{BELLE}:-2.2<Re(\tilde{d}_{\tau})<4.5 \times (10^{-17}\, e\,cm),
\end{eqnarray}
\begin{eqnarray}
\text{BELLE}:-2.5<Im(\tilde{d}_{\tau})<0.8 \times (10^{-17}\, e\,cm).
\end{eqnarray}

As a result, the magnetic and electric dipole moments of the tau lepton allow stringent testing for new physics beyond SM and have been studied in detail by Refs. \cite{d2,d3,d4,d5,d6,d7,d8,d9,d10,d11,d12,d13,d15,d16,d17,d18,d19,d20,d21,d22,d23,d24,pich}.

The advantage of the linear colliders with respect to the hadron colliders is in the general cleanliness of the events where two elementary particles, electron and positron beams, collide at high energy, and the high resolutions of the detector are made possible by the relatively low absolute rate of background events.
One of the most realistic linear colliders is the ILC. According to the LHC, thanks to the clean event environment, the ILC may be able to observe the smallest deviation from the SM estimates that point to new physics, discover new particles and make precise measurements of them. The ILC is planned to run at center-of-mass energies of 250, 350 and 500 GeV, with total integrated luminosities of 2000, 200 and 4000 fb$^{-1}$, respectively \cite{rr}. An increase of up to 1 TeV is also considered for center-of-mass energy, which we do not include in our calculations \cite{rr1}.

\section{CROSS SECTIONS AND SENSITIVITY ANALYSIS}

The effects of anomalous contributions arising from dimension-six operators and SM contributions as well as interference between new physics and the SM contribution is performed through the process $\gamma \gamma \rightarrow \tau \bar{\tau}\gamma$ at the ILC. For this purpose, firstly, in Fig. 1, the Feynman diagrams of the process $\gamma \gamma \rightarrow \tau \bar{\tau}\gamma$ are represented. As seen in Fig. 1, the total number of diagrams is 6. Here, the contribution to new physics comes from all diagrams. Incoming photons in the process $\gamma \gamma \rightarrow \tau \bar{\tau}\gamma$ are Compton backscattered photons. The spectrum of Compton backscattered photons is given as follows

\begin{eqnarray}
f_{\gamma/e}(y)=\frac{1}{g(\zeta)}\left[1-y+\frac{1}{1-y}-\frac{4y}{\zeta(1-y)}+\frac{4y^{2}}{\zeta^{2}(1-y)^{2}}\right]
\end{eqnarray}
where

\begin{eqnarray}
g(\zeta)=\left(1-\frac{4}{\zeta}-\frac{8}{\zeta^{2}}\right) log(\zeta+1)+\frac{1}{2}+\frac{8}{\zeta}-\frac{1}{2(\zeta+1)^{2}}
\end{eqnarray}
with

\begin{eqnarray}
y=\frac{E_{\gamma}}{E_{e}},\ \zeta=\frac{4E_{0}E_{e}}{M^{2}_{e}},\ y_{max}=\frac{\zeta}{1+\zeta}
\end{eqnarray}
where $E_{\gamma}$ represents the energy of the backscattered, $E_{0}$ and $E_{e}$ show energy of the incoming laser photon
and initial energy of the electron beam before Compton backscattering. Also, the maximum value of $y$ reaches 0.83 when
$\zeta=4.8$.

\begin{figure} [ht]
\includegraphics [width=12cm,height=6cm]{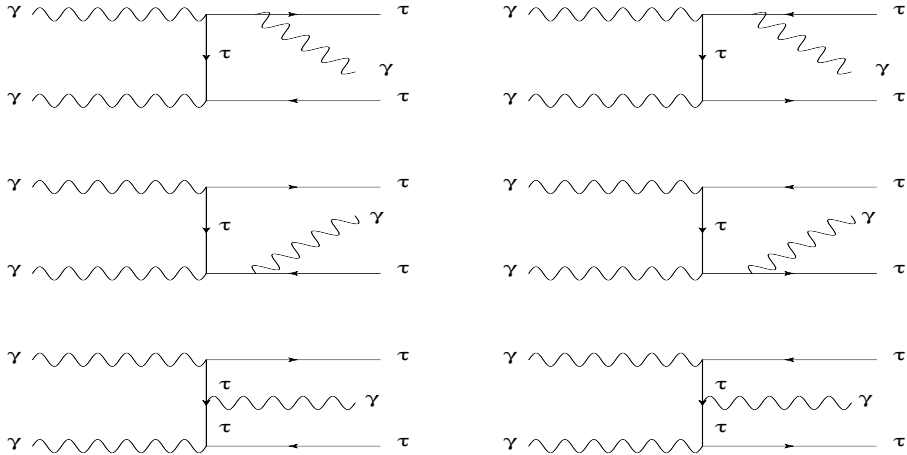}
\caption{Tree-level Feynman diagrams for the process $\gamma\gamma \rightarrow \tau \bar{\tau} \gamma$.
\label{fig1}}
\end{figure}

The total cross section of the process $\gamma \gamma \rightarrow \tau \bar{\tau}\gamma$ can be given by the following integration:

\begin{eqnarray}
d\sigma(e^{-}e^{+}\rightarrow \gamma \gamma \rightarrow \tau \bar{\tau}\gamma)=\int^{z_{max}}_{z_{min}} dz\,2z\, d\hat{\sigma}(\gamma \gamma \rightarrow \tau \bar{\tau}\gamma)\times \int^{y_{max}}_{z^{2}/y_{max}}\frac{dy}{y}f_{\gamma/e}(y)f_{\gamma/e}(z^{2}/y).
\end{eqnarray}
Here, $d\hat{\sigma}(\gamma \gamma \rightarrow \tau \bar{\tau}\gamma)$ represents the cross section of the process
and the center-of-mass energy of $e^{-}e^{+}$ system, $\sqrt{s}$, is related to the center-of-mass energy of $\gamma \gamma$ system, $\sqrt{\hat{s}}$ by $\hat{s}=z^{2}s$. The total cross section of process $\gamma \gamma \rightarrow \tau \bar{\tau}\gamma$ is an even function of $\tilde{\kappa}$ and a nonzero value of this parameter always has a constructive effect on the total cross section. Thus, contribution to the total cross section of $\tilde{\kappa}$ is proportional to $\tilde{\kappa}^{2}$ or higher order even power:

\begin{eqnarray}
\sigma(\tilde{\kappa})=\sigma^{3} \tilde{\kappa}^{6}+ \sigma^{2} \tilde{\kappa}^{4}+\sigma^{1} \tilde{\kappa}^{2}+\sigma_{SM}.
\end{eqnarray}
Besides, the effect of ${\kappa}$ parameter on the total cross section is given by:

\begin{eqnarray}
\sigma(\tilde{\kappa})=\sigma^{9} \kappa^{6}+ \sigma^{8} \kappa^{5}+\sigma^{7} \kappa^{4}+\sigma^{6} \kappa^{3}+\sigma^{5} \kappa^{2}+\sigma^{4} \kappa+\sigma_{SM}.
\end{eqnarray}
Here, $\sigma_{SM}$ is the contribution of the SM, $\sigma^{i}(i=1-9)$ are the anomalous contribution. Also, $\sigma^{9}$ and $\sigma^{3}$ coefficients are same \cite{rr2}. It can be seen that the total cross sections of the process $\gamma \gamma \rightarrow \tau \bar{\tau}\gamma$ are symmetric for the anomalous $\tilde{\kappa}$ coupling, it is nonsymmetric for ${\kappa}$. Thus, we anticipate that while the limits on the anomalous magnetic dipole moment are asymmetric, the limits on the electric dipole moment are symmetric.

We know that the high dimensional operators could affect $p_{T}$ distribution of the photon, specially at the region with a large $p_{T}$ values, which can be very useful to distinguish signal and background events. For this purpose, we use the following cuts set: $p^{\tau ,\bar{\tau},\gamma}_{T}$, $|\eta^{\tau, \bar{\tau},\gamma}|$, $\Delta R(\tau, \bar{\tau}), (\tau, \gamma) ,(\bar{\tau},\gamma)$ where $p_{T}$ is transverse momentum of the particles in the final state, $|\eta|$ is the pseudorapidity of the particles in the final state and $\Delta R$ is the the separation of the particles in the final state. We apply $p^{\tau ,\bar{\tau}}_{T}>15$ GeV, $p^{\gamma}_{T}>10$ GeV, $|\eta^{\tau, \bar{\tau},\gamma}|<2.5$  and $\Delta R(\tau, \bar{\tau}), (\tau, \gamma) ,(\bar{\tau},\gamma)>0.4$ with tagged Cut-1, four different values of $p^{\gamma}_{T}$ with tagged Cut-2, Cut-3, Cut-4, and Cut-5 changing according to center-of-mass energies. A summary of cuts set shows in Table I. For three center-of-mass energies, the total cross sections of the process $\gamma \gamma \rightarrow \tau \bar{\tau}\gamma$ as a function of the anomalous $\kappa$ and $\tilde{\kappa}$ couplings for kinematic
cuts described in Table I are given in Figs. 2-7. As can be seen from these figures, the changes of the total and the SM cross sections according to $\kappa$ and $\tilde{\kappa}$ couplings have similar characteristics. In addition, after each kinematic cut is applied, the cross sections decrease as expected. To take a closer look at these rates of change, Table II has been presented. In Table II, we give the total cross sections and the SM cross section of the process $\gamma \gamma \rightarrow \tau \bar{\tau}\gamma$ with respect to different kinematic cuts for $\kappa=\tilde{\kappa}=0.03$. As can be understood from Table II, the ratios arise from the total cross sections divided by the SM cross sections and increase after each applied kinematic cut. As the applied kinematic cuts increase, the SM cross section is suppressed, thus the signal becomes more apparent. Thus, we observe the total and SM cross sections of the process $\gamma \gamma \rightarrow \tau \bar{\tau}\gamma$ with Cut-5 of each center-of-mass energy.

\begin{table} [ht]
\caption{Descriptions of kinematic cuts used for analysis.}
\label{tab2}
\begin{tabular}{p{3cm}p{5cm}}
\hline \hline
Cuts & Definitions \\
\hline
Cut-1 & $p^{\tau ,\bar{\tau}}_{T}>15$ GeV $+$ $p_T^\gamma >10$ GeV
$+$ $|\eta^{\tau, \bar{\tau},\gamma}|<2.5$
$+$ $\Delta R(\tau, \bar{\tau}), (\tau, \gamma) ,(\bar{\tau},\gamma)>0.4$\\
Cut-2 & Same as in Cut-1, but for $p_T^\gamma >15$ GeV\\
Cut-3 & Same as in Cut-1, but for $p_T^\gamma >20$ GeV\\
Cut-4 & Same as in Cut-1, but for $p_T^\gamma >25$ GeV\\
Cut-5 & Same as in Cut-1, but for $p_T^\gamma >30$ GeV\\ \hline \hline

\end{tabular}
\end{table}

\begin{figure} [ht]
\includegraphics {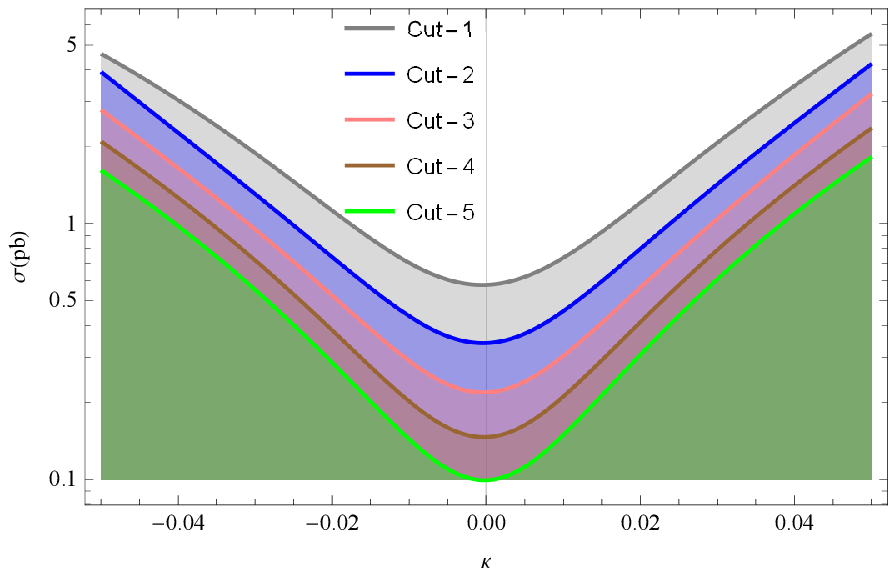}
\caption{The total cross sections of the process $\gamma\gamma \rightarrow \tau \bar{\tau} \gamma$ with $\sqrt{s}=250$ GeV beam as a function of the anomalous $\kappa$ couplings at five different kinematic cuts.
\label{fig1}}
\end{figure}

\begin{figure} [ht]
\includegraphics {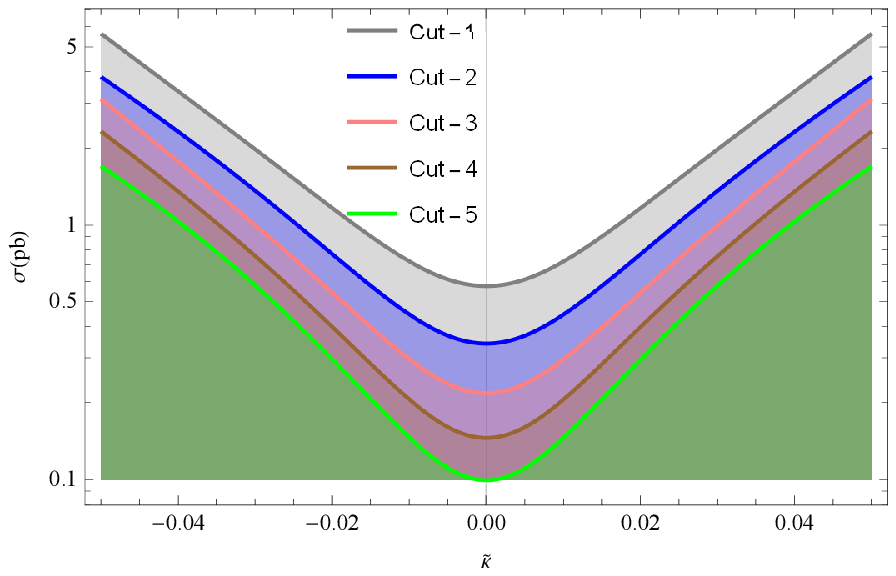}
\caption{Same as in Fig. 2, but for the anomalous $\tilde{\kappa}$ couplings.
\label{fig1}}
\end{figure}

\begin{figure} [ht]
\includegraphics {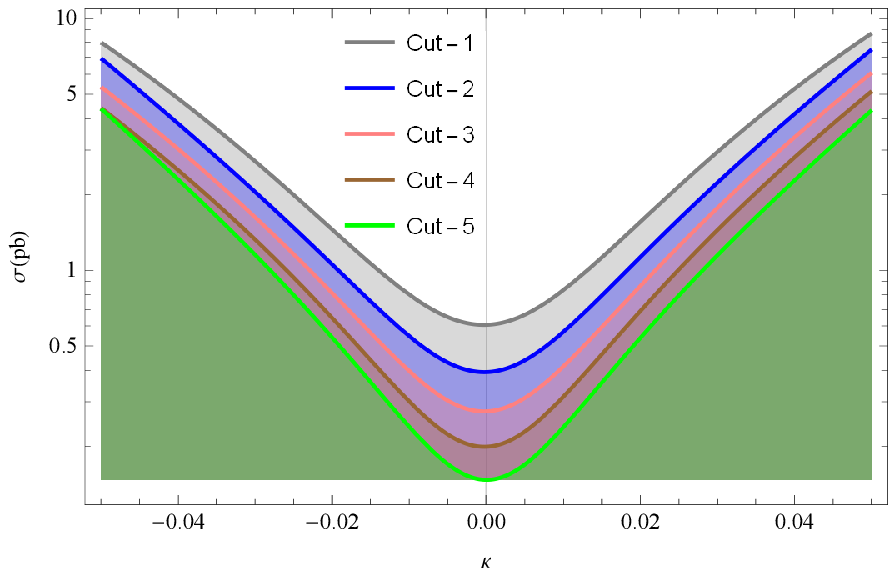}
\caption{The total cross sections of the process $\gamma\gamma \rightarrow \tau \bar{\tau} \gamma$ with $\sqrt{s}=350$ GeV beam as a function of the anomalous $\kappa$ couplings at five different kinematic cuts.
\label{fig1}}
\end{figure}

\begin{figure} [ht]
\includegraphics {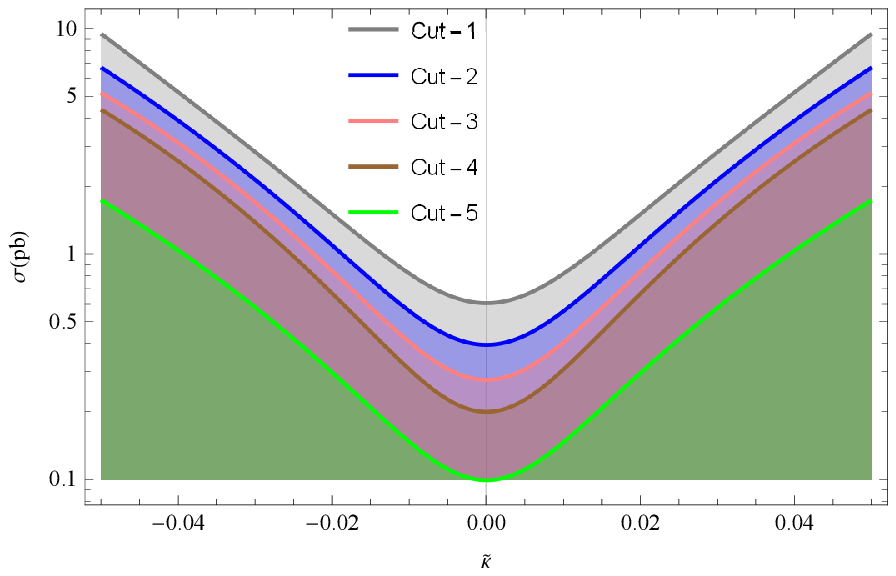}
\caption{Same as in Fig. 4, but for the anomalous $\tilde{\kappa}$ couplings.
\label{fig1}}
\end{figure}

\begin{figure} [ht]
\includegraphics {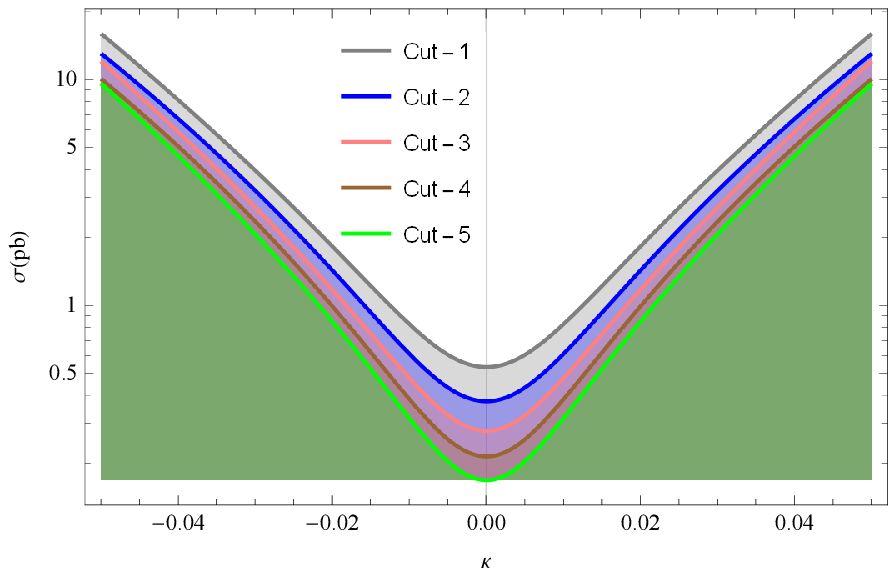}
\caption{The total cross sections of the process $\gamma\gamma \rightarrow \tau \bar{\tau} \gamma$ with $\sqrt{s}=500$ GeV beam as a function of the anomalous $\kappa$ couplings at five different kinematic cuts.
\label{fig1}}
\end{figure}

\begin{figure} [ht]
\includegraphics {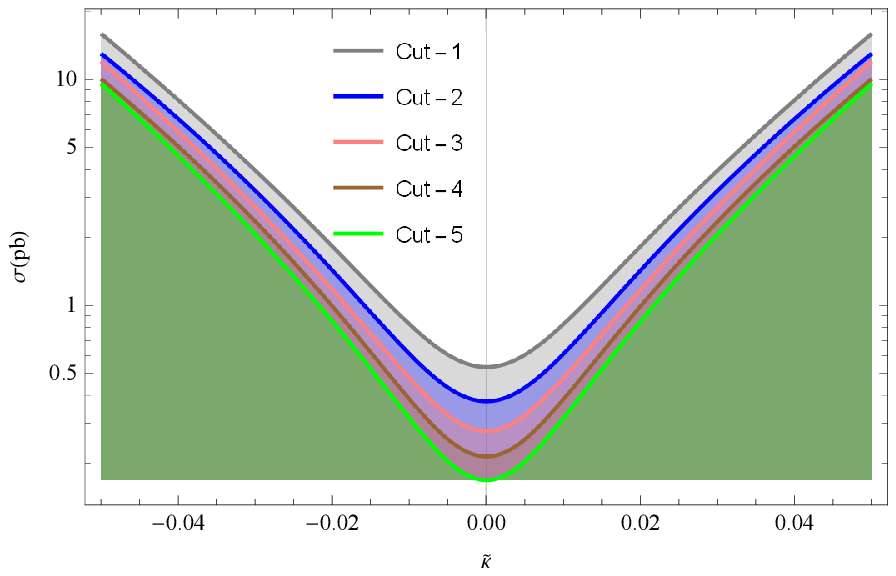}
\caption{Same as in Fig. 6, but for the anomalous $\tilde{\kappa}$ couplings.
\label{fig1}}
\end{figure}

\begin{table} [ht]
\caption{Total and SM cross section values and the ratios of total cross section to SM cross section on the anomalous $\kappa$ and $\tilde{\kappa}$ couplings for three different center-of-mass energies and five different cuts. Here, we assume that the anomalous $\kappa$ and $\tilde{\kappa}$ couplings are equal to $0.03$.}
\label{tab3}
\begin{ruledtabular}
\begin{tabular}{ccccccc}
& & & {$\kappa$}&  &$\tilde{\kappa}$ & \\
\hline
& \multirow{2}{*}{Cuts} & SM cross & Total cross & \multirow{2}{*}{Ratio} & Total cross & \multirow{2}{*}{Ratio}\\
Center-of-mass energy &  & sections & sections &  & sections & \\
 &  & (pb) & (pb) &  & (pb) &  \\
\hline \hline

 & Cut-1 & 0.574 & 2.076 & 3.617 & 1.982 & 3.453 \\
 & Cut-2 & 0.342 & 1.437 & 4.202 & 1.360 & 3.977 \\
$\sqrt{s}=250$ GeV & Cut-3 & 0.219 & 1.046 & 4.776 & 0.999 & 4.562 \\
& Cut-4 & 0.147 & 0.794 & 5.401 & 0.754 & 5.129 \\
 & Cut-5 & 0.099 & 0.608 & 6.141 & 0.579 & 5.848 \\ \hline

 & Cut-1 & 0.605 & 2.951 & 4.878 & 2.825 & 4.669 \\
 & Cut-2 & 0.395 & 2.227 & 5.638 & 2.134 & 5.403 \\
$\sqrt{s}=350$ GeV & Cut-3 & 0.275 & 1.767 & 6.425 & 1.691 & 6.149 \\
& Cut-4 & 0.199 & 1.445 & 7.261 & 1.379 & 6.930 \\
 & Cut-5 & 0.147 & 1.201 & 8.170 & 1.150 & 7.823 \\ \hline

 & Cut-1 & 0.536 & 4.107 & 7.662 & 3.955 & 7.379 \\
 & Cut-2 & 0.376 & 3.334 & 8.867 & 3.218 & 8.559 \\
$\sqrt{s}=500$ GeV & Cut-3 & 0.279 & 2.820 & 10.107 & 2.719 & 9.746 \\
& Cut-4 & 0.215 & 2.447 & 11.381 & 2.350 & 10.930 \\
 & Cut-5 & 0.168 & 2.145 & 12.768 & 2.064 & 12.286 \\
\end{tabular}
\end{ruledtabular}
\end{table}

Figs. 8 and 9 present the results for the total cross sections of the process $\gamma \gamma \rightarrow \tau \bar{\tau}\gamma$ at $\sqrt{s}=250, 350, 500$ GeV depending on $\kappa$ and $\tilde{\kappa}$ parameters. Here, we assume that only one of two anomalous couplings deviate from the SM at any given time. The total cross sections show a clear dependence according to the center-of-mass energy and the anomalous couplings. As can be seen from these figures, the deviation from the SM of the total cross sections including the anomalous couplings at $\sqrt{s}=500$ GeV is larger than the other center-of-mass energies. Thus, we expect that the obtained limits on the anomalous $\kappa$ and $\tilde{\kappa}$ couplings at $\sqrt{s}=500$ GeV are to be more restrictive than the limits at $\sqrt{s}=250,350$ GeV. Moreover, the effects of the anomalous $\kappa$ and $\tilde{\kappa}$ couplings on the total cross sections of the process $\gamma \gamma \rightarrow \tau \bar{\tau}\gamma$ are shown in Figs. 10-12.

\begin{figure}
\includegraphics{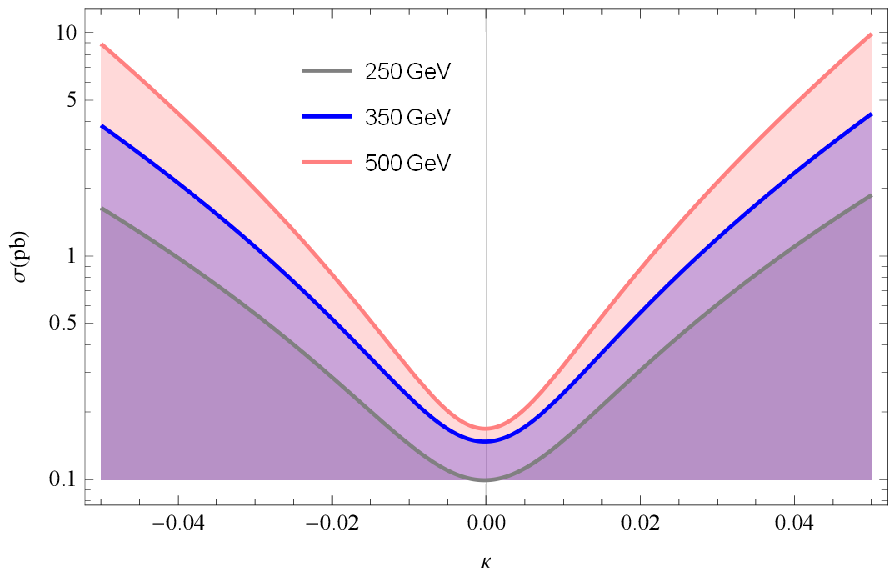}
\caption{The total cross sections of the process $\gamma \gamma \rightarrow \tau \bar{\tau}\gamma$ depending on the anomalous $\kappa$ coupling at $\sqrt{s}=250,350$ and $500$ GeV at the ILC.
\label{fig2}}
\end{figure}

\begin{figure}
\includegraphics{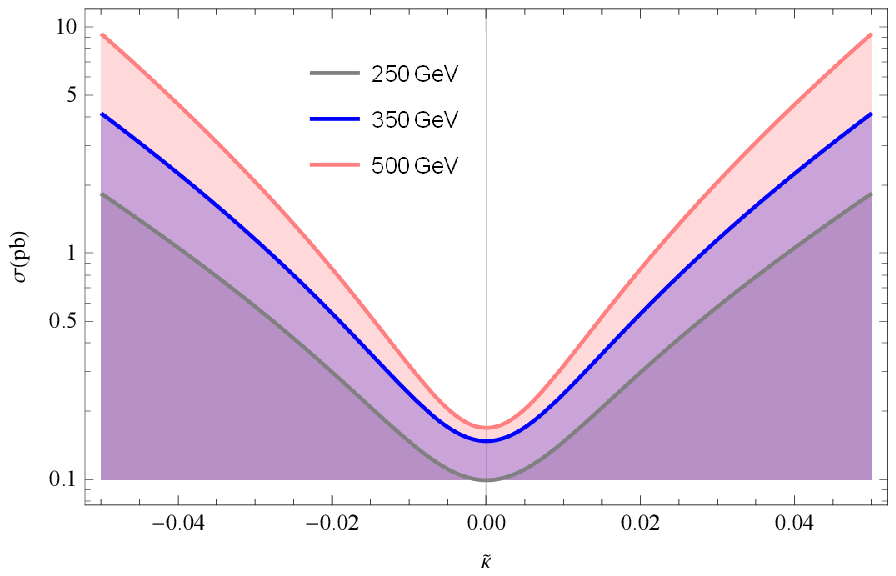}
\caption{Same as in Fig. 8, but for the anomalous $\tilde{\kappa}$ coupling.
\label{fig3}}
\end{figure}

\begin{figure}
\includegraphics{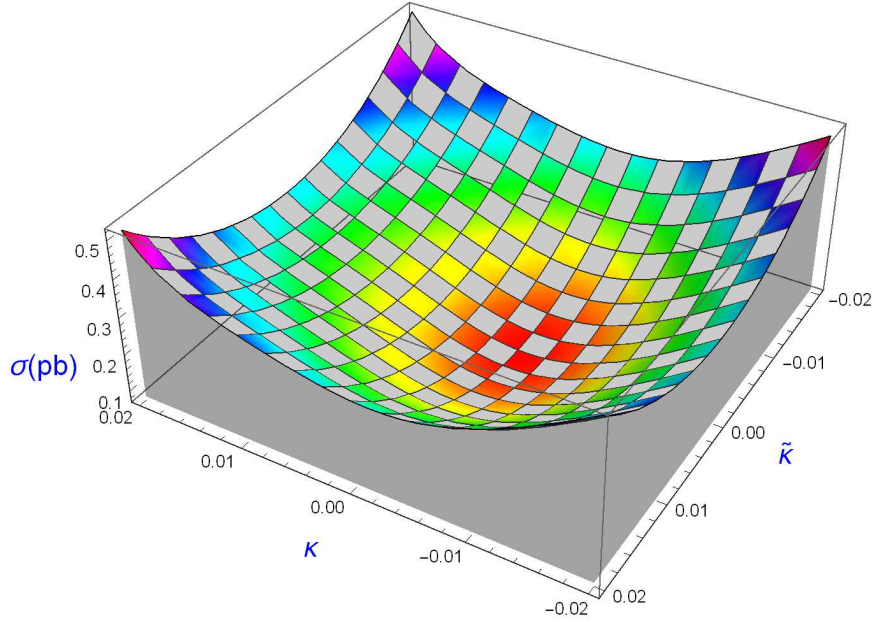}
\caption{The total cross sections of the process $\gamma \gamma \rightarrow \tau \bar{\tau}\gamma$ depending on the anomalous $\kappa$ and $\tilde{\kappa}$ couplings at $\sqrt{s}=250$ GeV at the ILC.
\label{fig4}}
\end{figure}

\begin{figure}
\includegraphics{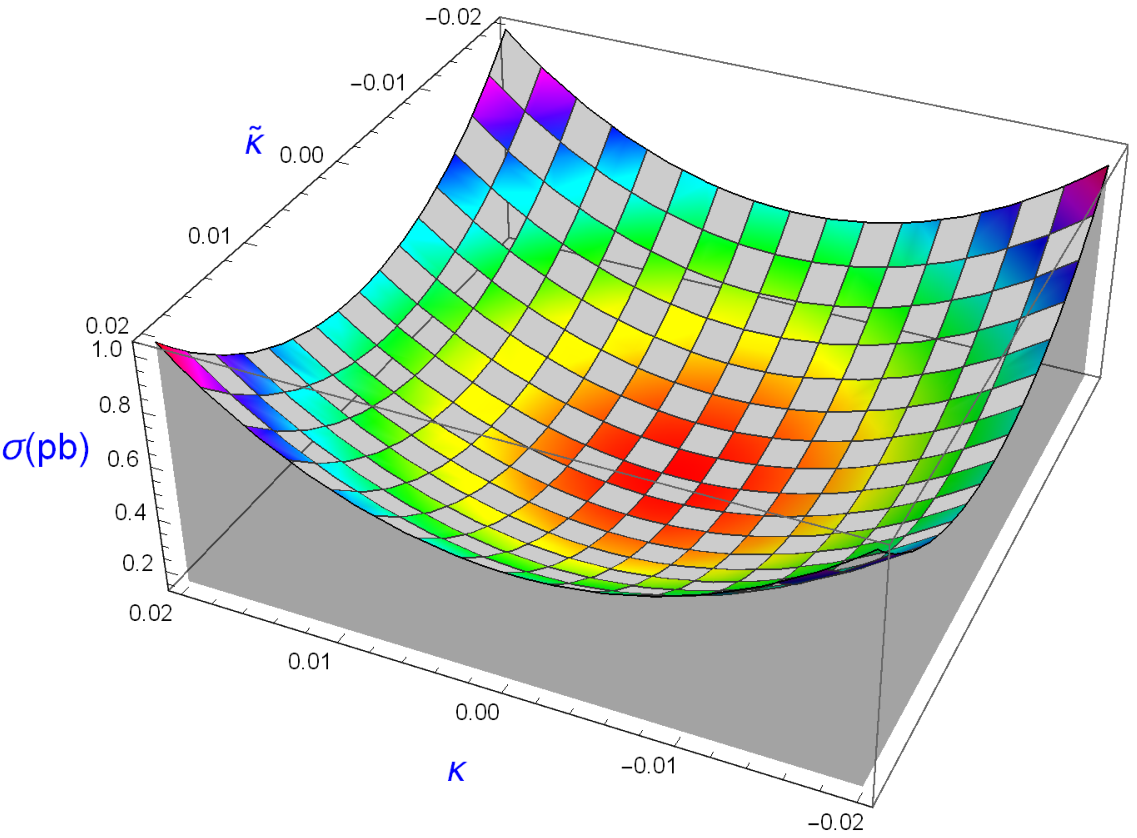}
\caption{Same as in Fig. 4, but for $\sqrt{s}=350$ GeV.
\label{fig5}}
\end{figure}

\begin{figure}
\includegraphics{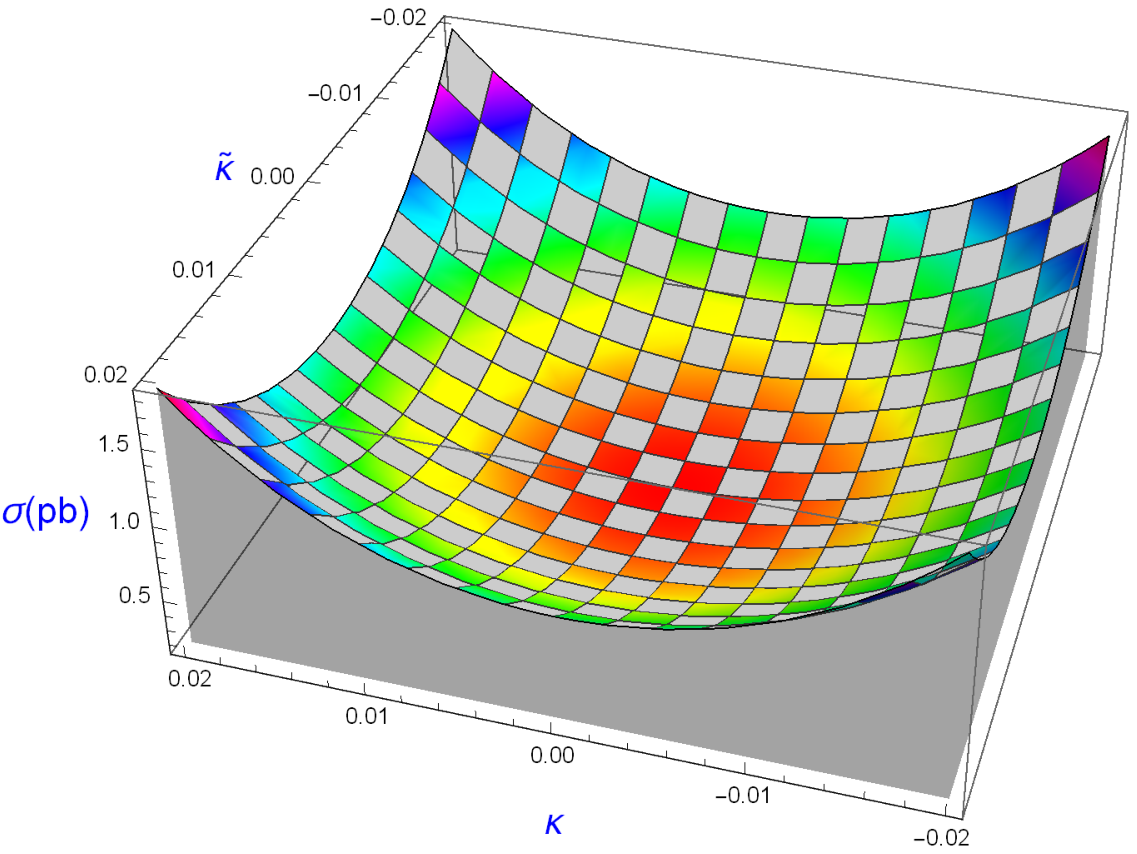}
\caption{Same as in Fig. 4, but for $\sqrt{s}=500$ GeV.
\label{fig6}}
\end{figure}

We use $\chi^{2}$ analysis with systematic error to study the sensitivities on the anomalous  $\tilde{a}_{\tau}$ and $\tilde{d}_{\tau}$ dipole moments of the $\tau$ lepton:

\begin{eqnarray}
\chi^{2}=\left(\frac{\sigma_{SM}-\sigma_{NP}}{\sigma_{SM}\delta}\right)^{2},
\end{eqnarray}

\begin{eqnarray}
\delta=\sqrt{\delta_{stat}^{2}+\delta_{sys}^{2}},
\end{eqnarray}

\begin{eqnarray}
N_{SM}=L_{int}\times BR \times \sigma_{SM},
\end{eqnarray}

\begin{eqnarray}
\chi^{2}=\left(\frac{\sigma_{SM}-\sigma_{NP}}{\sigma_{SM}\delta}\right)^{2},
\end{eqnarray}
where $BR$, $N_{SM}$ represent branching ratio and number of events. $L_{int}$ shows the integrated luminosity of the ILC. $\delta_{stat}$ and $\delta_{sys}$ are statistical and systematic uncertainties, respectively. The tau lepton is the only lepton that has the mass necessary
to disintegrate, most of the time in hadrons. In $17.8\%$ of the time, the tau lepton decays into an
electron and into two neutrinos; in another $17.4\%$ of the time, it decays in a muon and in
two neutrinos. In the remaining $64.8.8\%$ of the occasions, it decays in the form of hadrons
and a neutrino. In this work, we take into account pure leptonic and semileptonic decays for tau leptons in the final state.
Thus, we assume that in pure leptonic decays BR = 0.123 and in semileptonic decays BR = 0.46.

Systematic uncertainty takes place when the tau lepton is observed in colliders. Due to these uncertainties, tau identification efficiencies are calculated for the specific process, luminosity, and kinematic parameters. A detailed study is needed to achieve a realistic efficiency of a particular process. The systematic uncertainty is not exactly found in any ILC report for the process we are examining. However, there are many studies to probe the anomalous electromagnetic dipole moments of the tau lepton with systematic errors. The systematic uncertainty given at the LEP while investigating the electric and magnetic dipole moments of the tau lepton through the process $e^{-} e^{+} \rightarrow e^{-} e^{+} \tau^{-} \tau^{+}$ was between $4.3\%$ and $8.9\%$ \cite{de}.
Ref. \cite{d2} has investigated the tau anomalous magnetic moment with systematic errors of 0.1, 1 and $2\%$ at the prospect of future $e^{-}e^{+}$ colliders, such as the ILC, the CLIC, the FCC-ee and the CEPC. The sensitivity limits on the anomalous moments of the tau lepton via the process $e^{-} e^{+} \rightarrow e^{-} e^{+} \tau^{-} \tau^{+}$ at the CLIC were obtained by assuming up to $10\%$ systematic error \cite{d7}. The processes $p p \rightarrow p p \tau^{-} \tau^{+}$ and $p p \rightarrow p p \tau \bar{\nu}_{\tau}j$ were examined from $2\%$ to $7\%$ with systematic uncertainties at the LHC by Refs. \cite{d12,d13}. Thus, taking these studies into account, we will obtain limits on the magnetic and electric dipole moments of the tau lepton by $0,5,10\%$ systematic errors.

In Tables III-VIII, the limits obtained at $95\%$ confidence level on the anomalous  $\tilde{a}_{\tau}$ and $\tilde{d}_{\tau}$ dipole moments of the tau lepton via the process $\gamma \gamma \rightarrow \tau \bar{\tau}\gamma$ in the case of two decay channels at the ILC with $\delta_{sys}=0,5,10\%$ are represented.
The best ILC sensitivity limits on the anomalous  $\tilde{a}_{\tau}$ and $\tilde{d}_{\tau}$ dipole moments of the tau lepton might reach up to the order of magnitude $\mathcal{O}(10^{-4}-10^{-3})$ and $\mathcal{O}(10^{-18}-10^{-17})$, respectively. As can be seen in Table VIII, the best limits obtained on $\tilde{a}_{\tau}$ and $\tilde{d}_{\tau}$ are $-0.00082<\tilde{a}_{\tau}<0.00050$ and $|\tilde{{d}_{\tau}}|<3.5969 \times 10^{-18}$ $e\,cm$, respectively. Thus, our best limits on the anomalous $\tilde{a}_{\tau}$ and $\tilde{d}_{\tau}$ couplings improve much better than the experimental limits.

\begin{table}
\caption{For systematic errors of $0,5\%$ and $10\%$, the limits on the anomalous couplings at $\sqrt{s}=250$ GeV ILC through pure leptonic decay channel with integrated luminosities of $100, 500, 1000$ and $2000$ fb$^{-1}$.}
\begin{ruledtabular}
\begin{tabular}{ccccc}
$\delta_{sys}$& Luminosity($fb^{-1}$)&$\tilde{a}_{\tau}$ & $\vert \tilde{d}_{\tau} \vert (e\,cm) $  \\
\hline
&$100$ &(-$0.00375$, $ 0.00319$)  &$1.9185\times 10^{-17}$ \\
$0\%$&$500$ &(-$0.00260$, $0.00206$) &$1.2836\times 10^{-17}$ \\
&$1000$ &(-$0.00224$, $0.00169$) &$1.0795\times 10^{-17}$ \\
&$2000$ &(-$0.00193$, $0.00139$) &$9.0783\times 10^{-18}$ \\
\hline
&$100$ &(-$0.00520$, $ 0.00462$)  &$2.7167\times 10^{-17}$ \\
$5\%$&$500$ &(-$0.00493$, $0.00436$) &$2.5710\times 10^{-17}$ \\
&$1000$ &(-$0.00490$, $0.00433$) &$2.5510\times 10^{-17}$ \\
&$2000$ &(-$0.00488$, $0.00431$) &$2.5408\times 10^{-17}$ \\
\hline
&$100$ &(-$0.00689$, $ 0.00628$)  &$3.6445\times 10^{-17}$ \\
$10\%$&$500$ &(-$0.00678$, $0.00618$) &$3.5880\times 10^{-17}$ \\
&$1000$ &(-$0.00677$, $0.00617$) &$3.5807\times 10^{-17}$ \\
&$2000$ &(-$0.00676$, $0.00616$) &$3.5771\times 10^{-17}$ \\
\end{tabular}
\end{ruledtabular}
\end{table}

\begin{table}
\caption{Same as in Table III, but for semileptonic decay channel.}
\begin{ruledtabular}
\begin{tabular}{ccccc}
$\delta_{sys}$& Luminosity($fb^{-1}$)&$\tilde{a}_{\tau}$ & $\vert \tilde{d}_{\tau} \vert (e\,cm) $  \\
\hline
&$100$ &(-$0.00277$, $ 0.00223$)  &$1.3788\times 10^{-17}$ \\
$0\%$&$500$ &(-$0.00195$, $0.00141$) &$9.2228\times 10^{-18}$ \\
&$1000$ &(-$0.00169$, $0.00115$) &$7.7558\times 10^{-18}$ \\
&$2000$ &(-$0.00147$, $0.00093$) &$6.5221\times 10^{-18}$ \\
\hline
&$100$ &(-$0.00496$, $ 0.00439$)  &$2.5841\times 10^{-17}$ \\
$5\%$&$500$ &(-$0.00488$, $0.00431$) &$2.5415\times 10^{-17}$ \\
&$1000$ &(-$0.00487$, $0.00430$) &$2.5360\times 10^{-17}$ \\
&$2000$ &(-$0.00486$, $0.00430$) &$2.5332\times 10^{-17}$ \\
\hline
&$100$ &(-$0.00679$, $0.00619$)  &$3.5928\times 10^{-17}$ \\
$10\%$&$500$ &(-$0.00676$, $0.00616$) &$3.5773\times 10^{-17}$ \\
&$1000$ &(-$0.00676$, $0.00616$) &$3.5754\times 10^{-17}$ \\
&$2000$ &(-$0.00676$, $0.00616$) &$3.5744\times 10^{-17}$ \\
\end{tabular}
\end{ruledtabular}
\end{table}

\begin{table}
\caption{For systematic errors of $0,5\%$ and $10\%$, the limits on the anomalous couplings at $\sqrt{s}=350$ GeV ILC through pure leptonic decay channel with integrated luminosities of $10, 50, 100$ and $200$ fb$^{-1}$.}
\begin{ruledtabular}
\begin{tabular}{ccccc}
$\delta_{sys}$& Luminosity($fb^{-1}$)&$\tilde{a}_{\tau}$ & $\vert \tilde{d}_{\tau} \vert (e\,cm) $  \\
\hline
&$10$ &(-$0.00516$, $0.00469$)  &$2.7332\times 10^{-17}$ \\
$0\%$&$50$ &(-$0.00353$, $0.00308$) &$ 1.8310\times 10^{-17}$ \\
&$100$ &(-$0.00301$, $0.00256$) &$1.5403\times 10^{-17}$ \\
&$200$ &(-$0.00256$, $0.00212$) &$1.2956\times 10^{-17}$ \\
\hline
&$10$ &(-$0.00564$, $0.00517$) &$2.9983\times 10^{-17}$ \\
$5\%$&$50$ &(-$0.00466$, $ 0.00420$) &$2.4574\times 10^{-17}$ \\
&$100$ &(-$0.00448$, $0.00402$) &$2.3577\times 10^{-17}$ \\
&$200$ &(-$0.00438$, $0.00392$) &$ 2.3026\times 10^{-17}$ \\
\hline
&$10$ &(-$0.00661$, $0.00612$) &$3.5304\times 10^{-17}$ \\
$10\%$& $50$ &(-$0.00610$, $0.00562$) &$3.2491\times 10^{-17}$ \\
&$100$ &(-$0.00602$, $0.00554$) &$3.2082\times 10^{-17}$ \\
&$200$ &(-$0.00598$, $0.00551$) &$3.1872\times 10^{-17}$ \\
\end{tabular}
\end{ruledtabular}
\end{table}

\begin{table}
\caption{Same as in Table V, but for semileptonic decay channel.}
\begin{ruledtabular}
\begin{tabular}{ccccc}
$\delta_{sys}$& Luminosity($fb^{-1}$)&$\tilde{a}_{\tau}$ & $\vert \tilde{d}_{\tau} \vert (e\,cm) $  \\
\hline
&$10$ &(-$0.00377$, $0.00332$)  &$1.9665\times 10^{-17}$ \\
$0\%$&$50$ &(-$0.00260$, $0.00216$) &$1.3162\times 10^{-17}$ \\
&$100$ &(-$0.00223$, $0.00178$) &$1.1071\times 10^{-17}$ \\
&$200$ &(-$0.00191$, $0.00147$) &$9.3110\times 10^{-18}$ \\
\hline
&$10$ &(-$0.00477$, $0.00431$) &$ 2.5173\times 10^{-17}$ \\
$5\%$&$50$ &(-$0.00439$, $0.00393$) &$2.3064\times 10^{-17}$ \\
&$100$ &(-$0.00433$, $0.00387$) &$2.2755\times 10^{-17}$ \\
&$200$ &(-$0.00430$, $0.00385$) &$2.2595\times 10^{-17}$ \\
\hline
&$10$ &(-$0.00614$, $0.00566$) &$3.2753\times 10^{-17}$ \\
$10\%$& $50$ &(-$0.00599$, $0.00551$) &$3.1885\times 10^{-17}$ \\
&$100$ &(-$0.00597$, $0.00549$) &$3.1772\times 10^{-17}$ \\
&$200$ &(-$0.00596$, $0.00548$) &$ 3.1772\times 10^{-17}$ \\
\end{tabular}
\end{ruledtabular}
\end{table}

\begin{table}
\caption{For systematic errors of $0,5\%$ and $10\%$, the limits on the anomalous couplings at $\sqrt{s}=500$ GeV ILC through pure leptonic decay channel with integrated luminosities of $100, 1000, 2000$ and $4000$ fb$^{-1}$.}
\begin{ruledtabular}
\begin{tabular}{ccccc}
$\delta_{sys}$& Luminosity($fb^{-1}$)&$\tilde{a}_{\tau}$ & $\vert \tilde{d}_{\tau} \vert (e\,cm) $  \\
\hline
&$100$ &(-$0.00242$, $0.00210$)  &$1.2577\times 10^{-17}$ \\
$0\%$&$1000$ &(-$0.00144$, $0.00112$) &$7.0790\times 10^{-18}$ \\
&$2000$ &(-$0.00124$, $0.00092$) &$5.9534\times 10^{-18}$ \\
&$4000$ &(-$0.00107$, $0.00075$) &$5.0066\times 10^{-18}$ \\
\hline
&$100$ &(-$0.00372$, $0.00339$)  &$1.9782\times 10^{-17}$ \\
$5\%$&$1000$ &(-$0.00358$, $0.00325$) &$1.9021\times 10^{-17}$ \\
&$2000$ &(-$0.00357$, $0.00324$) &$1.8976\times 10^{-17}$ \\
&$4000$ &(-$0.00357$, $0.00324$) &$1.8953\times 10^{-17}$ \\
\hline
&$100$ &(-$0.00502$, $0.00468$)  &$2.7012\times 10^{-17}$ \\
$10\%$&$1000$ &(-$0.00497$, $0.00463$) &$ 2.6730\times 10^{-17}$ \\
&$2000$ &(-$0.00497$, $0.00463$) &$2.6714\times 10^{-17}$ \\
&$4000$ &(-$0.00497$, $0.00463$) &$2.6706\times 10^{-17}$ \\
\end{tabular}
\end{ruledtabular}
\end{table}

\begin{table}
\caption{Same as in Table VII, but for semileptonic decay channel.}
\begin{ruledtabular}
\begin{tabular}{ccccc}
$\delta_{sys}$& Luminosity($fb^{-1}$)&$\tilde{a}_{\tau}$ & $\vert \tilde{d}_{\tau} \vert (e\,cm) $  \\
\hline
&$100$ &(-$0.00179$, $0.00147$)  &$9.0408\times 10^{-18}$ \\
$0\%$&$1000$ &(-$0.00108$, $0.00076$) &$ 5.0862\times 10^{-18}$ \\
&$2000$ &(-$0.00094$, $0.00062$) &$4.2773\times 10^{-18}$ \\
&$4000$ &(-$0.00082$, $0.00050$) &$3.5969\times 10^{-18}$ \\
\hline
&$100$ &(-$0.00361$, $0.00328$)  &$1.9169\times 10^{-17}$ \\
$5\%$&$1000$ &(-$0.00357$, $0.00324$) &$1.8955\times 10^{-17}$ \\
&$2000$ &(-$0.00357$, $0.00324$) &$1.8943\times 10^{-17}$ \\
&$4000$ &(-$0.00357$, $0.00324$) &$1.8937\times 10^{-17}$ \\
\hline
&$100$ &(-$0.00498$, $0.00464$)  &$2.6782\times 10^{-17}$ \\
$10\%$&$1000$ &(-$0.00497$, $0.00463$) &$ 2.6706\times 10^{-17}$ \\
&$2000$ &(-$0.00497$, $0.00463$) &$2.6702\times 10^{-17}$ \\
&$4000$ &(-$0.00496$, $0.00463$) &$2.6700\times 10^{-17}$ \\
\end{tabular}
\end{ruledtabular}
\end{table}

We observe that the sensitivity obtained on the $\tilde{d}_{\tau}$ coupling from $\gamma\gamma$ collisions at the 250 GeV ILC are at the same order with the experimental limits, while the sensitivity on $\tilde{a}_{\tau}$ is expected to improve up to two orders of magnitude with respect to experimental results.
Also, for semi leptonic decays and without systematic error, our sensitivities on the anomalous couplings for the process $\gamma \gamma \rightarrow \tau \bar{\tau}\gamma$ with $\sqrt{s}=250$ GeV and $L_{int}=2000$ fb$^{-1}$ can set more stringent up to 1.5 times better than the best sensitivity derived from $\tau \tau \gamma$ production at the ILC with $\sqrt{s}=350$ GeV and $L_{int}=200$ fb$^{-1}$.

Tables III-VIII show that the limits with increasing the luminosity on the anomalous couplings do not increase proportionately to the luminosity due to the systematic error considered here.  The reason of this situation is the systematic error which is much
bigger than the statistical error. If the systematic error is improved, we expect better limits on the couplings. For example, our best limits on the anomalous couplings for the process $\gamma \gamma \rightarrow \tau \bar{\tau}\gamma$ with $\sqrt{s}=500$ GeV, $L_{int}=4000$ fb$^{-1}$ and $\delta_{sys}=0\%$
can be improved up to 4 times for $\tilde{a}_{\tau}$ and 9 times for $\tilde{d}_{\tau}$ according to case with $\delta_{sys}=10\%$.

Finally, the contours for the anomalous couplings for the process $\gamma \gamma \rightarrow \tau \bar{\tau}\gamma$ at the ILC for various integrated luminosities and center-of-mass energies are presented in Figs. 13-15. As we can see from these figures, the improvement in the sensitivity on the anomalous couplings is achieved by increasing to higher center-of-mass energies and luminosities.

\begin{figure}
\includegraphics{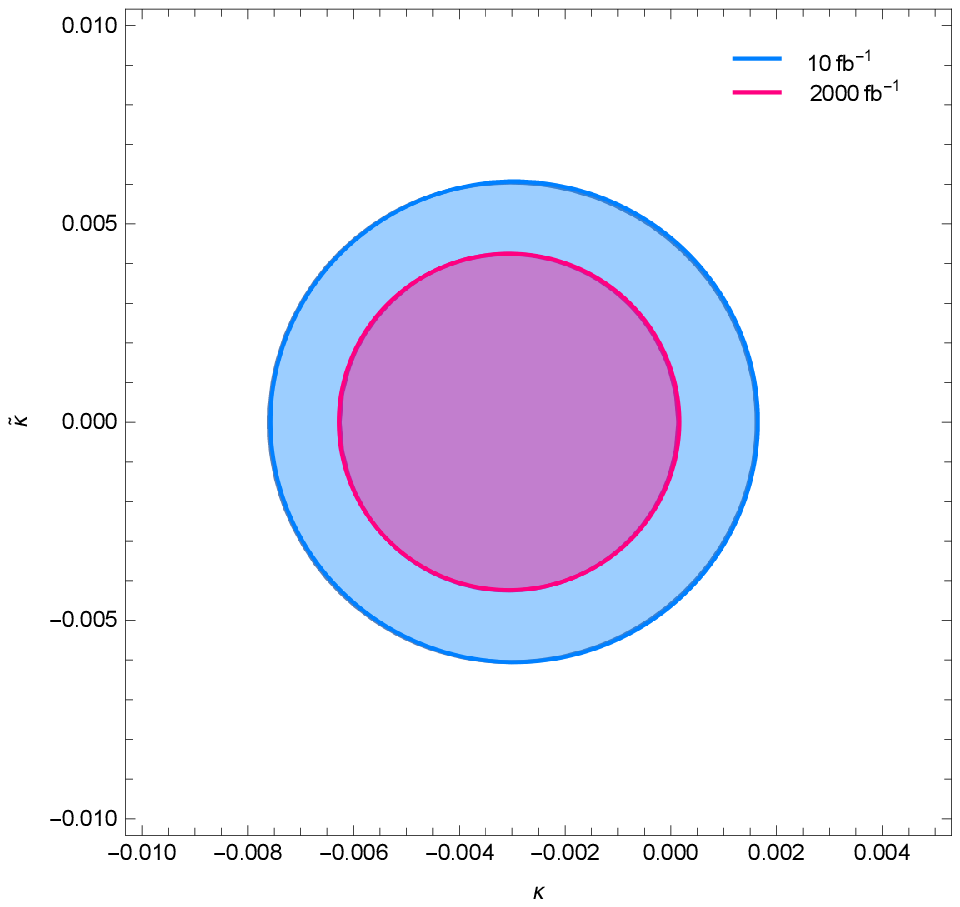}
\caption{For semi leptonic decay channel, $95\%$ confidence level contours for anomalous $\kappa$ and $\tilde{\kappa}$ couplings for the process $\gamma \gamma \rightarrow \tau \bar{\tau}\gamma$ with $\sqrt{s}=250$ GeV and $L_{int}=10, 2000$ fb$^{-1}$.
\label{fig4}}
\end{figure}

\begin{figure}
\includegraphics{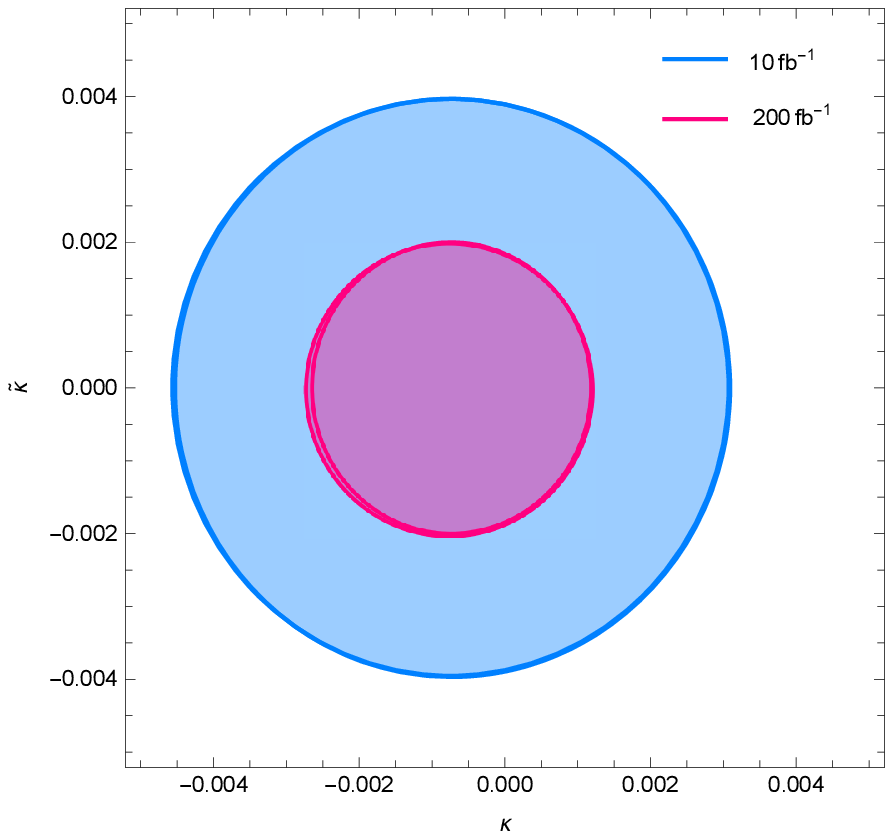}
\caption{The same as Figure 7 but for $\sqrt{s}=350$ GeV and $L_{int}=10, 200$ fb$^{-1}$.
\label{fig4}}
\end{figure}

\begin{figure}
\includegraphics{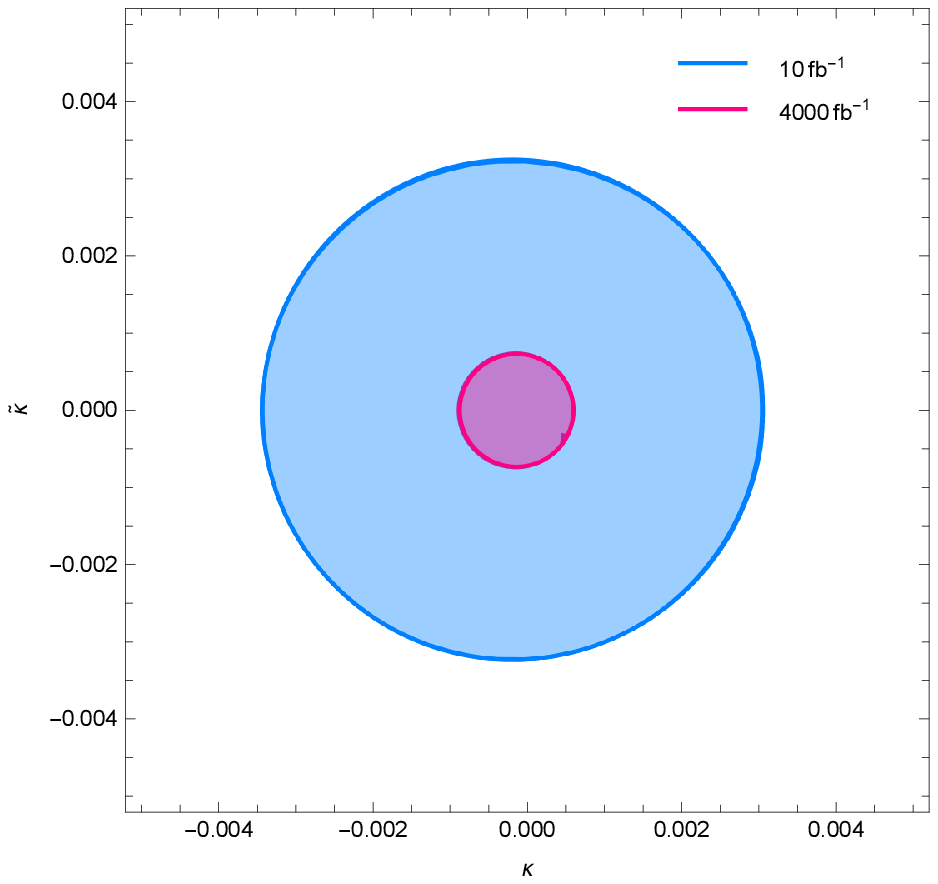}
\caption{The same as Figure 7 but for $\sqrt{s}=500$ GeV and $L_{int}=10, 4000$ fb$^{-1}$.
\label{fig4}}
\end{figure}

\section{Conclusions}

It is of great interest to examine and suggest mechanisms model independent to study the magnetic and electric dipole moments of the tau lepton with the processes examined in the colliders. The $\tau \bar{\tau} \gamma$ coupling between the tau lepton and the photon needs to be studied precisely. Since non-standard $\tau \bar{\tau} \gamma$ couplings defined via effective Lagrangian have dimension-six, they have very strong energy dependencies. Thus, the anomalous cross sections including $\tau \bar{\tau} \gamma$ vertex have a higher energy than the SM cross section.
In this case, a possible deviation from the SM cross section of any process involving $\tau \bar{\tau} \gamma$ coupling may be a sign of the existence of the new physics. For this purpose, we have investigated on the phenomenological aspects of the anomalous $\tau \bar{\tau} \gamma$ couplings with
the process $\gamma \gamma \rightarrow \tau \bar{\tau}\gamma$ at the ILC. The total cross section and the limit analysis are
analyzed in regard to the the anomalous $\tilde{a}_{\tau}$ and $\tilde{d}_{\tau}$ dipole moments which cause the possible deviations from the SM.
Here, a cutflow is created according to cuts set to achieve optimized limits. The ratio to SM cross section of total cross section including the the anomalous $\tilde{a}_{\tau}$ and $\tilde{d}_{\tau}$ parameters has determined and the contributions of the kinematic cuts to the signal have examined. Also, for the anomalous $\tilde{a}_{\tau}$ and $\tilde{d}_{\tau}$ parameters, the limits at $95\%$ confidence level by using $\chi^{2}$ test are obtained.  We find that the ILC with $\sqrt{s}=500$ GeV and 4000 fb$^{-1}$ give the best limits on the all anomalous coupling parameters.

The process $\gamma \gamma \rightarrow \tau \bar{\tau} \gamma$ has some advantages. The anomalous $\tau \bar{\tau}\gamma$ couplings can be analyzed via the process $e^{-}e^{+} \rightarrow \tau \bar{\tau}\gamma$ at the linear colliders. This process receives contributions from both the anomalous $\tau \bar{\tau}\gamma$ and $\tau \bar{\tau}Z$ couplings. Nevertheless, the process $\gamma \gamma \rightarrow \tau \bar{\tau}\gamma$ isolate $\tau \bar{\tau}\gamma$ coupling, and thus $\tau \bar{\tau}\gamma$ and $\tau \bar{\tau}Z$ couplings may be investigated separately.  Moreover, the single photon in the final state has the advantage of being identifiable with high efficiency and purity. For this reason, the selection criteria used for the analysis enables examining for events with single-photon characteristics.  Finally, $\gamma \gamma$ collisions in the lepton colliders may be effective efficient $\tau$ identification due to clean final state when compared to hadron colliders.

Consequently, we emphasize that the sensitivities obtained on the anomalous $\tilde{a}_{\tau}$ and $\tilde{d}_{\tau}$ couplings in our work are better than
the sensitivity of the experimental limits. $\gamma \gamma$ collisions at the ILC to investigate the anomalous $\tilde{a}_{\tau}$ and $\tilde{d}_{\tau}$ couplings via the process $\gamma \gamma \rightarrow \tau \bar{\tau}\gamma$ are quite suitable for investigating the anomalous $\tilde{a}_{\tau}$ and $\tilde{d}_{\tau}$ dipole moments of the $\tau$ lepton.

\pagebreak

\newpage

\end{document}